\documentclass[a4paper,12pt,oneside]{article}
\usepackage[warn]{mathtext}
\usepackage{amsmath}
\usepackage[T2A]{fontenc}
\usepackage{latexsym}
\usepackage[english]{babel}
\usepackage{graphics}
\usepackage{indentfirst}

\title{Parametric (quasi-Cherenkov) cooperative radiation produced by electron bunches in natural or photonic crystals}
\author{{\sc S. V. Anishchenko}${}^{1}$\thanks{E--mail: {\tt
      sanishchenko@mail.ru}},{\sc V. G. Baryshevsky}${}^{1}$\thanks{E--mail: {\tt
      bar@inp.bsu.by}}\\
{\small\em ${}^{1}$ Nuclear Problems Institute, Bobruiskaya 11, Minsk 220030, Belarus}}

\date{}
\begin{document}
\maketitle
\begin{abstract}
We use numerical modeling to study the features of parametric
(quasi-Cherenkov) cooperative radiation arising when an electron bunch
passes through a crystal (natural or artificial) under the
conditions of dynamical diffraction of electromagnetic waves in
the presence of shot noise.
It is shown that in both Laue and Bragg diffraction cases,
parametric radiation consists of two strong pulses: one
emitted at small angles  with respect to the particle velocity
direction and the other emitted at large angles to it.
Under Bragg diffraction conditions, the intensity of parametric radiation emitted at small
angles to the particle velocity direction reaches  saturation at
sufficiently smaller number of particles than the intensity of parametric radiation emitted at large angles.
Under Laue diffraction conditions, every pulse contains two strong
peaks, which are associated with the emission of electromagnetic
waves at the front and back ends of the bunch. The presence of
noise causes a chaotic signal in the interval between the two
peaks.

\end{abstract}

\section{Introduction}

The generation of short pulses of electromagnetic radiation is a
primary challenge of modern physics. They find applications in
studying molecular dynamics in biological objects and  charge
transfer in nanoelectronic devices, diagnostics of dense plasma
and radar detection of fast moving objects.

The advances in the generation of short pulses of electromagnetic
radiation in infrared, visible, and ultraviolet ranges of
wavelengths  are traditionally associated with the development of
quantum electronic devices - lasers. Radiation in lasers  is
generated via induced  emission of photons by bound electrons.

Electrovacuum devices, operating in a cooperative regime
~\cite{Bonifacio1990,Korovin2006},  have recently become
considered as an alternative to short-pulse lasers, whose active
medium is formed by electrons bound in atoms and molecules. These
are free electron lasers, cyclotron-resonance masers, and
Cherenkov radiators, whose active medium is formed by electron
bunches propagating in complex electrodynamical structures
(undulators, corrugated waveguides and others). The feature of the
cooperative operation regime lies in the fact that the radiation
power scales as the squared number of particles in the bunch.
This allows calling this regime "superradiance"  by
analogy with the phenomenon predicted by Dicke in quantum
electronics \cite{Bonifacio1990}. 
It should be noted, that the electrons moving in FELs are initially homogeneously distributed in the phase space. As a result, bremsstrahlung starts from sponteneous emission. This is true even if the bunch length is much smaller than the radiation wave length. In contrast to bremsstrahlung, Cherenkov (quasi-Cherenkov) radiation starts from coherent emission when such a short-length bunch is injected into a slow-wave structure, i. e. the radiation power is proportional to the squared number of particles. The question arises whether this dependence holds when the bunch length increases.

%


%
This paper considers coperative radiation emitted by electron bunches
when charged particles pass through crystals (natural or
artificial) under the conditions of dynamical diffraction of
elctromagnetic waves.
Note that a detailed analysis of the features of spontaneous
radiation of electrons passing through crystals in both frequency
\cite{BFU} and time \cite{AnishchenkoBaryshevskyGurinovich2012}
domains has been carried out before. This radiation, emitted at
both large and small angles with respect to the direction of
electron motion, is called the parametric (quasi-Cherenkov) radiation. The problems
of amplification of induced parametric radiation
 have also been thoroughly studied
in the literature \cite{BaryshevskyFeranchuk1984}, and the
threshold current densities providing lasing in crystals have been
calculated \cite{Baryshevsky2012}.

This paper is arranged as follows: In the beginning, a nonlinear
theory of interaction of relativistic charged particles and the
electromagnetic field in crystals is set forth, followed by the
results of numerical calculations of the parametric radiation
pulse. Then, the dependence of the radiation
intensity on the electron bunch length and the geometrical
parameters of the system is considered. The appendix outlines the
algorithm used in the simulation.  The feature of the algorithm is
that it is based on the particle-in-cell method, which enables
studying kinetic phenomena. Let us note that in most of the
existing codes (see, e.g. \cite{Antonsen,Ginzburg}) used
for simulating the interaction of charged particles and a
synchronous wave, the motion of charged particles is considered
within the framework of the hydrodynamic approximation.

\section{Nonlinear theory of cooperative radiation}

A theoretical analysis of parametric radiation can be performed
only by means of a  self-consistent solution of a nonlinear set of
the Vlasov--Maxwell equations:
\begin{equation}
\label{Vlasov}
 \frac{\partial f}{\partial t}+\vec v\frac{\partial f}{\partial \vec r}+q_e(\vec E+\vec v\times\vec H)\frac{\partial f}{\partial \vec p}=0,
\end{equation}
\begin{equation}
\label{Maxwell}
\begin{split}
 &\nabla\times\vec E=-\frac{1}{c}\frac{\partial\vec H}{\partial t},\\
 &\nabla\times\vec H=\frac{1}{c}\frac{\partial\vec D}{\partial t}+\frac{4\pi}{c}\vec j,\text{ }\vec j=q_e\int\vec vfd^3\vec p,\\
 &\nabla\cdot\vec D=4\pi\rho,\text{ }\rho=q_e\int fd^3\vec p,\\
 &\nabla\cdot\vec H=0,\\
\end{split}
\end{equation}
describing the electron motion in the electric $\vec E$ and
magnetic $\vec H$ fields.
Here $f(\vec r,\vec p,t)$ is the particle distribution
function over the coordinates $\vec r$ and momenta $\vec p$,
while $\vec j(\vec r, t)$ and $\rho(\vec r, t)$ are the current
and charge densities, respectively. Since the crystal is a
periodic linear medium with frequency dispersion,  the Fourier
transform of the displacement current $\vec D(\vec r,\omega)$
relates to the electric field $\vec E(\vec r,\omega)$ as $D(\vec
r,
\omega)=\big(1+\chi_0(\omega)+\sum\limits_{\vec\tau}2\chi_{\vec\tau}(\omega)\cos(\vec\tau\vec
r)\big)\vec E(\vec r, \omega)$, where the summation is made over
all reciprocal lattice vectors.
Because the dielectric susceptibilities in natural crystals and in
grid photonic crystals are inversely proportional to the frequency
\cite{BaryshevskyGurinovich2006}:
$\chi_{0,\vec\tau}(\omega)=\Omega_{0,\vec\tau}^2/\omega^2$,
Maxwell's equations \eqref{Maxwell} can be reduced to the equation
of the form:
\begin{equation}
\label{Maxwell2}
\frac{1}{c^2}\frac{\partial^2\vec E}{\partial t^2}+\nabla(\nabla\cdot\vec E)-\Delta\vec E+\frac{\Omega_0^2}{c^2}\vec E+\sum\limits_{\vec\tau}2\frac{\Omega_{\vec\tau}^2}{c^2}\cos(\vec\tau\vec r)\vec E=-\frac{4\pi}{c^2}\frac{\partial\vec j}{\partial t}.
\end{equation}

We shall further be interested in the case when two strong waves
are excited in the crystal: the forward wave and the diffracted
wave (the so-called two-wave diffraction case). The forward wave
(its wave   vector is denoted by $\vec k$) is emitted at small
angles with respect to the particle velocity, while the diffracted
one, having the wave vector $\vec k_\tau=\vec k+\vec\tau$,  is
emitted at large angles to it (Fig.~1). Under the conditions of
dynamical diffraction, the following relation is fulfilled: $\vec
k_\tau^2\approx \vec k^2\approx\omega^2/c^2$.

\begin{figure}[ht]
\begin{center}
       \resizebox{85mm}{!}{\includegraphics{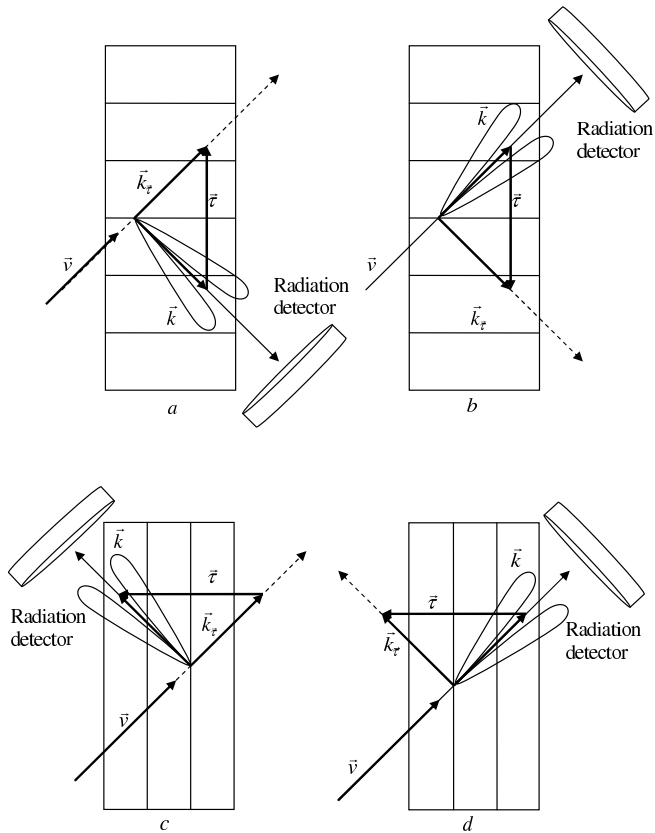}}\\
\caption{Schemes of parametric radiation in Laue (a, b) and Bragg (c, d)
 geometries.}
\end{center}
\end{figure}

Let us perform the following  simplifications: First,  we shall
neglect the longitudinal ($\nabla\cdot\vec D\to0$) fields of the
bunch, which is appropriate when the value of the Langmuir
oscillation frequency $\Omega_b$ of the bunch is less than the
values of $\Omega_{0,\tau}$. In this case, the Coulomb forces will
not have an appreciable effect on  dielectric properties of the
system. Second, we shall seek for the electric field $\vec E$
using the method of slowly varying amplitudes. Third, we shall
assume that a transversally infinite bunch executes
one-dimensional motion along the $OZ$-axis (this is achieved by
inducing a strong axial magnetic field in the system).

Under the conditions of two-wave diffraction, the field $\vec E$
can be presented as a sum
\begin{equation}
\label{ElectricField}
 \vec E=\vec e_0E_0(z,t)e^{i(\vec k\vec r-\omega t)}+\vec e_\tau E_\tau(z,t) e^{i(\vec k_\tau\vec r-\omega t)},
\end{equation}
where the amplitudes of the forward $E_0$ and diffracted $E_\tau$
waves are slowly varying variables. This means that for the distances
comparable with the wavelength and the  times comparable with the
oscillation period, the values of  $E_0$ and~$E_\tau$ remain
practically the same. Substituting \eqref{ElectricField} into
\eqref{Vlasov} and  \eqref{Maxwell2} and then averaging them over
the length $l=2j\lambda=2j\pi/k_z$, where $j$ is a natural number,
we obtain
\begin{equation}
\label{Vlasov3}
 \frac{\partial f}{\partial t}+v_z\frac{\partial f}{\partial z}+2q_e\text{Re}\Big(E_0e^{i(k_zz-\omega t)}\Big)\frac{\partial f}{\partial p_z}=0,
\end{equation}
\begin{equation}
\label{Maxwell3}
\begin{split}
&\frac{1}{c}\frac{\partial E_0}{\partial t}+\gamma_0\frac{\partial E_0}{\partial z}+\frac{i\Omega_0^2}{2\omega c}E_0+\frac{i\Omega_\tau^2}{2\omega c}E_\tau=-\frac{2\pi}{c}\int_{z-l/2}^{z+l/2} e_{0z}j_ze^{i(\omega t-k_zz)}dz/l,\gamma_0=k_z/k,\\
&\frac{1}{c}\frac{\partial E_\tau}{\partial t}+\gamma_\tau\frac{\partial E_\tau}{\partial z}+\frac{i\Omega_0^2}{2\omega c}E_\tau+\frac{i\Omega_\tau^2}{2\omega c}E_0=-\frac{2\pi}{c}\int_{z-l/2}^{z+l/2} e_{\tau z}j_ze^{i(\omega t-k_{\tau z}z)}dz/l,\gamma_\tau=k_{\tau z}/k.\\
\end{split}
\end{equation}

Now let us complete the set of equations \eqref{Vlasov3} and
\eqref{Maxwell3} with boundary conditions (the initial conditions
are reduced to the condition that all values of the fields equal
zero at $t=0$): in the plane $z=0$, let us specify the time
dependence of function $f$ and set the field $E_0$  to zero. In
the case of Bragg diffraction, the boundary condition imposed on
the diffracted wave is reduced to the condition that the field
$E_\tau$ equals zero at $z=L$, while in the case of Laue
diffraction, it equals zero at $z=0$.

The difference between the two diffraction schemes is not merely
kinematic, but radical. Under Bragg diffraction conditions, there
is a synchronous wave moving against the electrons of the beam,
which gives rise to the  internal feedback and absolute
instability.
In Laue diffraction geometry, a backward wave is absent, and as a
result absolute instability does not evolve. 
It may
seem that electromagnetic radiation is not generated. However,
fluctuations, which always occur in real beams, are amplified when
the beam enters the  crystal (due to convective instability,
excited in the beam).

In analyzing multiparametric problems, to which the problem of
parametric cooperative radiation refers, it is convenient to write
equations \eqref{Vlasov3} and \eqref{Maxwell3} in a dimensionless
form. This procedure enabled transferring the calculation results
from one set of wavelengths and beam energies to another. The
analysis shows that this procedure is easy to perform in the case
when the electrons of the beam are ultrarelativistic (the Lorentz
factor $\gamma\gg1$).

Let us note that the kinetic equation  \eqref{Vlasov} is
equivalent to the system of relativistic equations of motion
 for the momenta $\vec p_j$  and coordinates $\vec r_j$ of
 particles, since $$f=\sum_j\delta(\vec r-\vec r_j)\delta(\vec p-\vec p_j)$$
 (the lower index $j$ runs over all beam's electrons  inside the crystal).
When $\gamma\gg1$, these equations are convenient to write with
the phases of charged particles ~$\phi_j=\vec k_0\vec r_j-\omega
t$ substituted for independent variables. The substitution of
variables $\omega t\theta^2\to t$, $\omega L\theta^2/c\to L$,
$mc\omega\theta E_{0,\tau}/q_e\to E_{0,\tau}$ ($\theta$ is the
angle between the particle velocity and the wave vector $\vec k$)
then gives

\begin{equation}
\label{Sim1}
 \frac{d^2\phi_j}{dt^2}=2\Big(-2\frac{d\phi_j}{dt}-1\Big)^{3/2}\text{Re}\Big(E_0e^{i\phi_j}\Big),
\end{equation}
\begin{equation}
\label{Sim2}
\begin{split}
&\frac{\partial E_0}{\partial t}+\gamma_0\frac{\partial E_0}{\partial z}+\frac{i\chi_0}{2\theta^2}E_0+\frac{i\chi_\tau}{2\theta^2}E_\tau=-\sum_j\frac{\chi_{bj}}{2\theta^2} \frac{e^{-i\phi_j}}{N_l},\\
&\frac{\partial E_\tau}{\partial t}+\gamma_\tau\frac{\partial E_\tau}{\partial z}+\frac{i\chi_0}{2\theta^2}E_\tau+\frac{i\chi_\tau}{2\theta^2}E_0=0.\\
\end{split}
\end{equation}
Here the quantity $\chi_{bj}=-4\pi q_e^2n_j/m\omega^2$ is
determined at the moment when the $j$th particle enters the
system, $n_j$ is the corresponding  electron density, and $N_l$ is
the number of particles over the length $l$.
The set of equations with boundary conditions contains four
independent parameters: $\chi_{0,\tau,b}/\theta^2$, $\omega L\theta^2/c$ that define
the geometry of the system. In addition to these parameters, we
need to specify the beam profile. Let
\begin{equation}
 \chi_b=\chi_{b0}(1-e^{-ct/L_f})e^{\Theta(ct-L_b)(ct-L_b)/L_f},
\end{equation}
where the rise time $L_f/c$ is further assumed to be equal
to $0.1L_b/c$ ($L_b$ is the bunch length and
 $\Theta$ is the Heaviside function).

\section{Simulation results}
 The characteristic feature of  cooperative pulses is the
 peculiar dependence of the peak power on the number of particles $N_b$ in the
 bunch. When the particles are small in number,  the radiation power monotonically increases until saturation is achieved.

\begin{figure}[ht]
\begin{center}
       \resizebox{75mm}{!}{\includegraphics{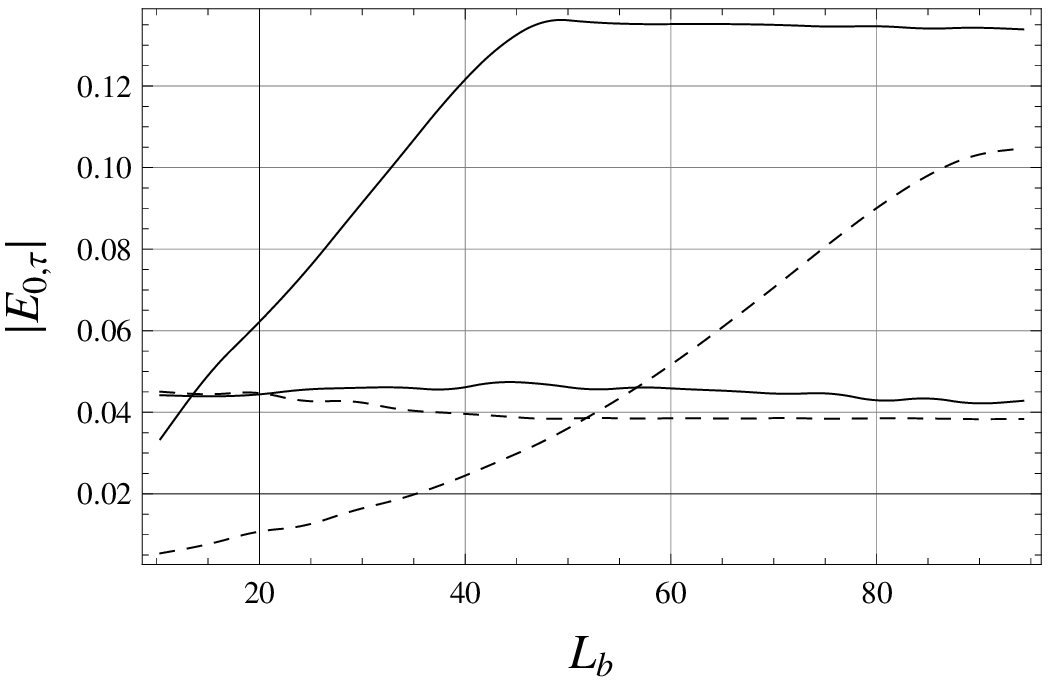}} \resizebox{75mm}{!}{\includegraphics{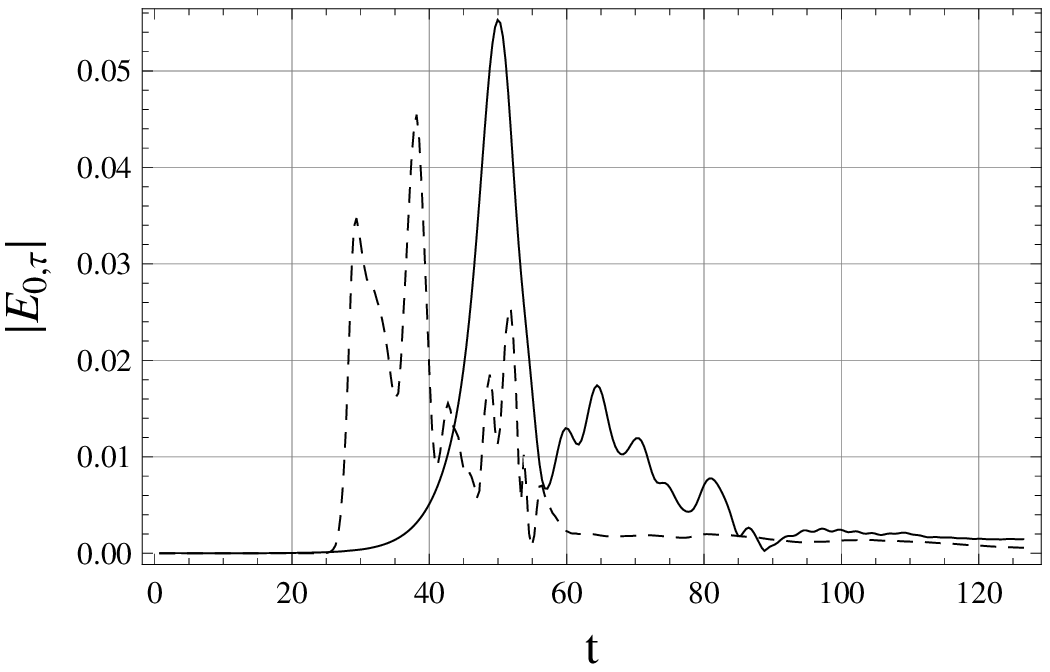}}\\
\caption{Parametric radiation under Bragg diffraction
conditions. Left: peak values of the fields $|E_{0,\tau}|$ plotted
as a function of the bunch length. Two upper curves are
plotted for radiation at large angels, while the two lower ones -
for radiation at small angles. Solid curves illustrate the case
when $\gamma_\tau=-1$; dashed curve - the case when
$\gamma_\tau=-0.5$. Right: the example of a cooperative pulse
[$\gamma_\tau=-1.0$]. Solid curve refers to a diffracted wave,
dashed curve --  to a direct one. }
\end{center}
\end{figure}

\begin{figure}[ht]
\begin{center}
\resizebox{75mm}{!}{\includegraphics{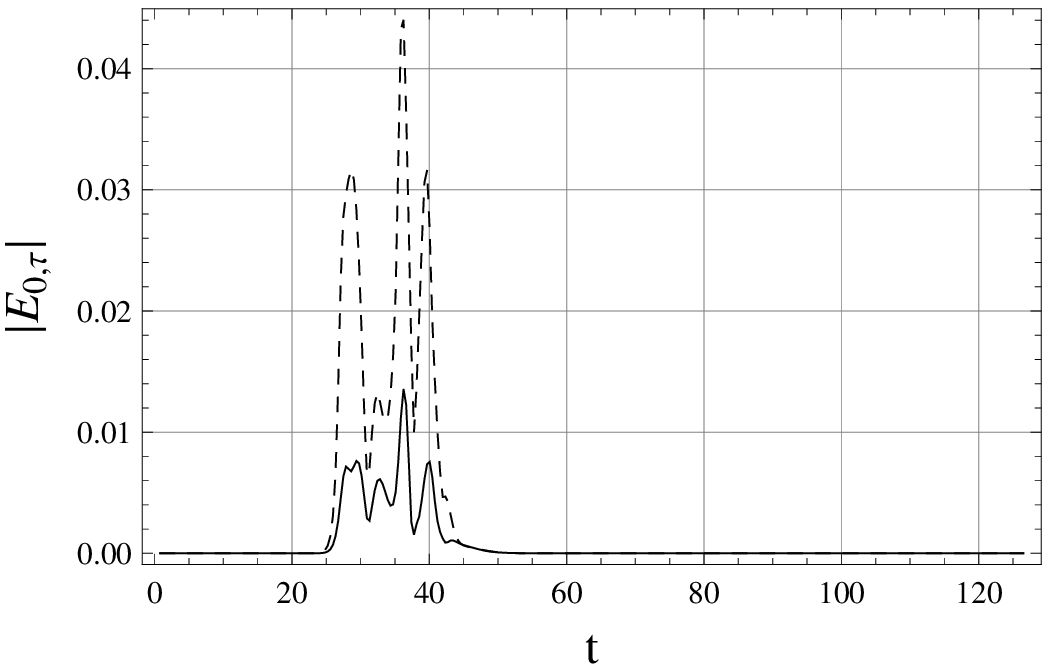}} \resizebox{75mm}{!}{\includegraphics{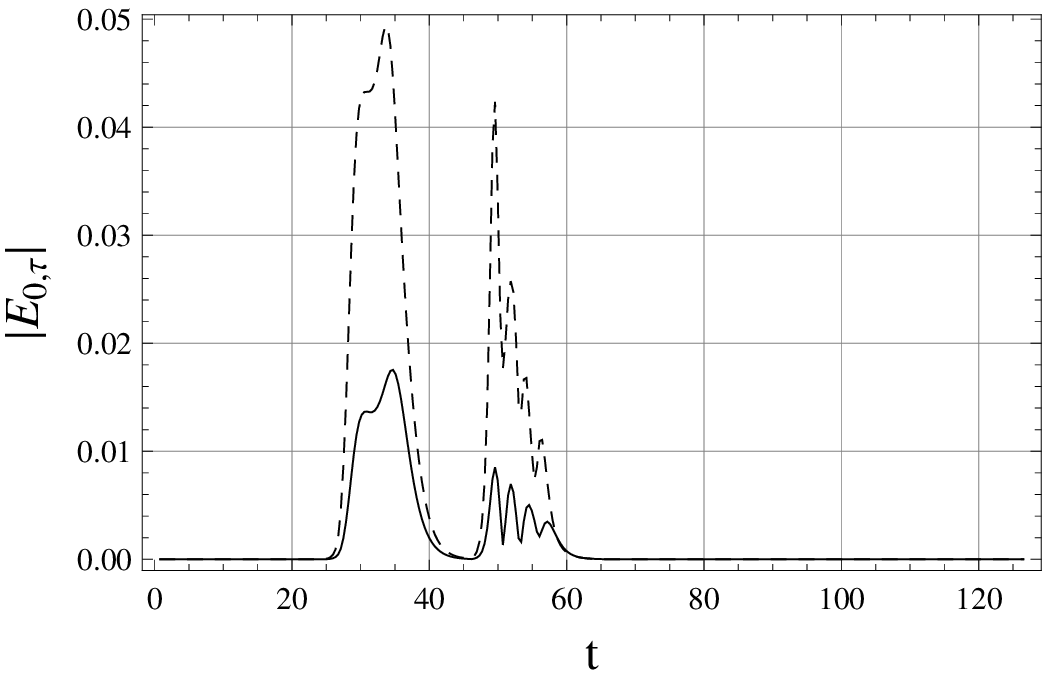}}\\
\caption{Parametric radiation under Laue diffraction
conditions. Solid curve is plotted for radiation at large angels,
while dashed curve -- for radiation at small angles.
 [$\gamma_\tau=-0.5$]. Left plot: $L_b=12.4$. Right plot: $L_b=24.9$. }

\end{center}
\end{figure}

\begin{figure}[ht]
\begin{center}
\resizebox{75mm}{!}{\includegraphics{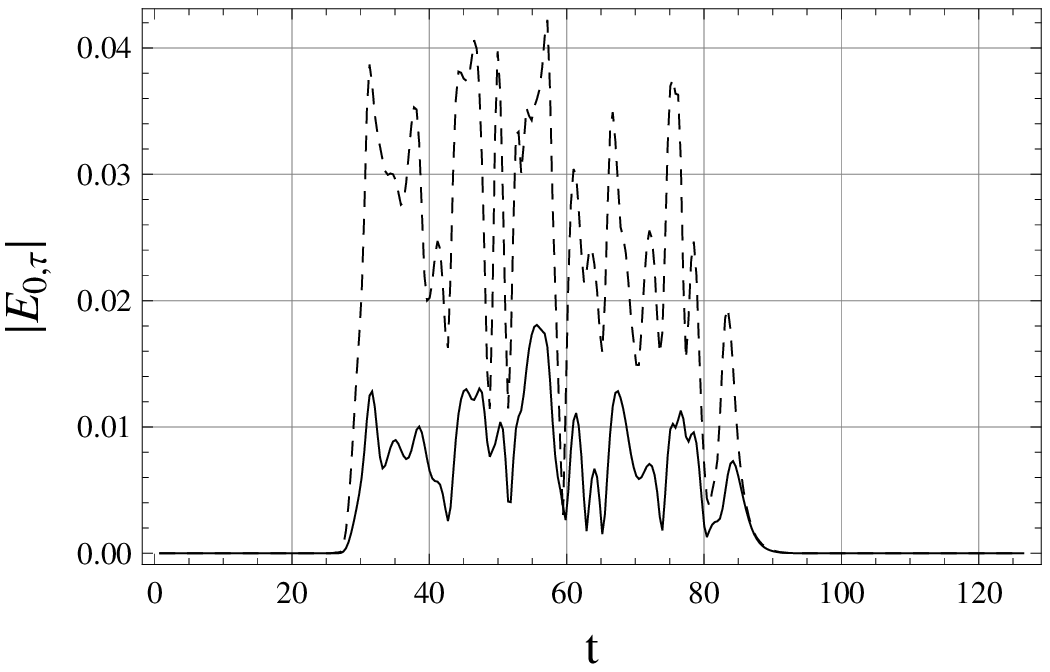}} \resizebox{75mm}{!}{\includegraphics{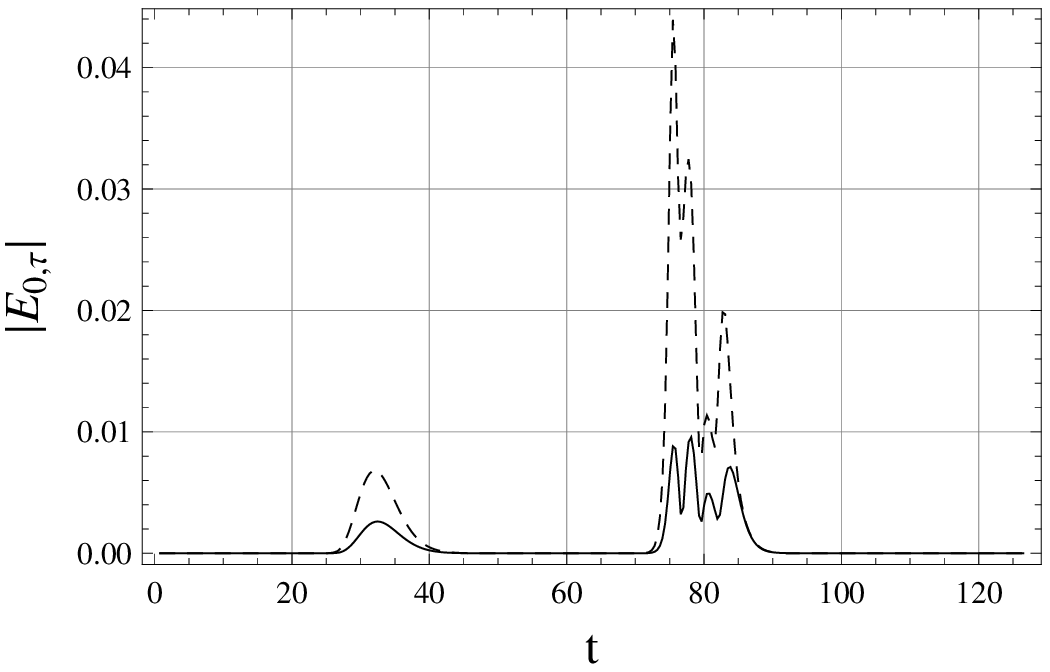}}\\
\caption{Pulse of cooperative radiation under Laue diffraction
conditions in the presence of noise (left) and without it (right).
Solid curve shows radiation at large angles, dashed curve -
radiation at small angles. The amplitude of white noise  ratio to
the average current density is assumed to be
~$j_{rand0}/j_{av}=0.01$. Computation is made for the case when
the bunch length $L_b=49.8$.}
\end{center}
\end{figure}

Let us see now how the dynamical diffraction of electromagnetic
waves affects  the cooperative radiation in crystals.
Does parametric radiation possess the above described features
inherent to other types of cooperative radiation? How the
Bragg diffraction case is different from the case of Laue
diffraction?

We shall assume that $\gamma=5$, $\chi_0=\chi_\tau=0.06$,
$\gamma_0=0.98$ and specify the system of units, where
 $c=1$, $\omega\theta^2=1$, $mc\omega\theta/q_e=1$ and $L=24.9$.

Let us start our consideration  with  Bragg diffraction.
The results of computation (Fig. 2) show that the
radiation emitted at small angles reaches saturation  at
significantly smaller lengths of the bunch than that emitted at
large angles.
The transition from a one-dimensional distributed feedback
($\gamma_\tau=-1$) to a two-dimensional one ($\gamma_\tau\neq-1$)
hardly causes  any change in the intensity of radiation emitted at
small angles to the particle velocity.

Under Laue diffraction conditions, the pulse of parametric
radiation has a more complicated structure due to the convective
character of instability: in the absence of noise, the generation
occurs only at the ends of the bunch of charged
particles. As a result, the cooperative pulse possess a
two-peak structure (Fig. 3).  The presence of noise leads to an
appreciable change in the pulse form: the interval between the two
pulses is filled with a chaotic signal (Fig.4).


\section{Conclusion}
This paper studies the features of parametric cooperative
radiation emitted at both large and small angles to the particle
velocity direction.  Both  Laue and Bragg cases are considered. It
is shown that under Bragg diffraction conditions, the intensity of
parametric radiation emitted at small angles reaches saturation at significantly smaller
number of particles in the bunch than
the intensity of radiation emitted at large angles.

Under Laue diffraction conditions, the pulses of parametric
radiation, emitted at both large and small angles to the particle
velocity, possess a two-peak structure; each of the peaks is
associated with  the front or back  end  of the bunch. The
presence of noise leads to an appreciable change in the pulse
form: the interval between the two pulses is filled with a chaotic
signal.

\appendix
\section{Particle-in-cell method}

The set of equations \eqref{Vlasov3} and \eqref{Maxwell3} was
solved using the particle-in-cell method, which is widely used in
plasma physics~\cite{Verboncoeur,SveshnikovJakunin1989}. This
method implies that the solution of the kinetic equation  is
modeled using a large number of macroparticles moving along the
characteristics of the kinetic equation. The current and charge
densities are calculated from particle velocities and positions
and are further used for computations of  the electric field on a
space-time mesh. The mesh values of the field are interpolated to
the macroparticle locations; then the forces acting on
macroparticles are calculated. The approach described here is
close to the method described by I.~J.~Morey and C.~K.~Birdsall in
\cite{MoreyBirdsall1989}, which was used for travelling wave tube
modeling.

Let us introduce a spatial  $\omega_z=\{z_n=n\Delta
z,n=0,1,...,n_{max},n_{max}\Delta z=L\}$ and a time
$\omega_t=\{t_s=s\Delta t,s=0,1,...\}$ mesh. Specify an implicit
finite-difference scheme of field equations  \eqref{Maxwell3} with
second order accuracy in time and coordinate:
\begin{equation}
\label{Maxwell4}
\begin{split}
&\text{at }  n=0: \text{ }E_{0n}^{s+1}=0,\\
&\frac{E_{\tau n}^{s+1}-E_{\tau n}^{s}}{c\Delta t}=-\gamma_\tau\frac{-3E_{\tau n}^{s+1/2}+4E_{\tau n+1}^{s+1/2}-E_{\tau n+2}^{s+1/2}}{2\Delta z}-\frac{i\chi_0}{2}E_{\tau n}^{s+1/2}-\frac{i\chi_\tau}{2}E_{0 n}^{s+1/2},\\
&\text{at }  n=n_{max}:\text{ }E_{\tau n}^{s+1}=0,\\
&\frac{E_{0n}^{s+1}-E_{0n}^{s}}{c\Delta t}=-\gamma_0\frac{3E_{0n}^{s+1/2}-4E_{0n-1}^{s+1/2}+E_{0n-2}^{s+1/2}}{2\Delta z}-\frac{i\chi_0}{2}E_{0 n}^{s+1/2}-\frac{i\chi_\tau}{2}E_{\tau n}^{s+1/2}-J_{0n}^{s+1/2},\\
&\text{at }  0<n<n_{max}:\\
&\frac{E_{0n}^{s+1}-E_{0n}^{s}}{c\Delta t}=-\gamma_0\frac{E_{0n+1}^{s+1/2}-E_{0n-1}^{s+1/2}}{2\Delta z}-\frac{i\chi_0}{2}E_{0 n}^{s+1/2}-\frac{i\chi_\tau}{2}E_{\tau n}^{s+1/2}-J_{0n}^{s+1/2},\\
&\frac{E_{\tau n}^{s+1}-E_{\tau n}^{s}}{c\Delta t}=-\gamma_\tau\frac{E_{\tau n+1}^{s+1/2}-E_{\tau n-1}^{s+1/2}}{2\Delta z}-\frac{i\chi_0}{2}E_{\tau n}^{s+1/2}-\frac{i\chi_\tau}{2}E_{0 n}^{s+1/2}.\\
\end{split}
\end{equation}

Let us define the source  $J_{0n}^{s+1/2}$ on the right-hand side
of
 \eqref{Maxwell4}, using the formula
\begin{equation}
\label{Source}
\begin{split}
J_{0n}^{s+1/2}=\frac{2\pi\sin\theta e^{i\omega (t+\Delta t/2)}}{cl}\Big(\sum_jQ_jv_{zj}^{s+1/2}\frac{z_{n+1}-z_j^{s+1/2}}{\Delta z}e^{-ik_zz_j^{s+1/2}}\\
+\sum_jQ_jv_{zj}^{s+1/2}\frac{z_j^{s+1/2}-z_{n-1}}{\Delta z}e^{-ik_zz_j^{s+1/2}}\Big).\\
\end{split}
\end{equation}
The contributions to each node come from the particles
concentrated in the domain   $z_{n-1}\le z_j<z_{n+1}$. Summation
in the first and second terms is made over all particles in the
domains $z_n\le z_j^{s+1/2}<z_{n+1}$ and $z_{n-1}\le
z_j^{s+1/2}<z_n$, respectively. The weighting factors
$\frac{z_{n+1}-z_j^{s+1/2}}{\Delta z}$ and
$\frac{z_j^{s+1/2}-z_{n-1}}{\Delta z}$ are responsible for linear
interpolation of the contributions to the node with number $n$
that come from each particle.

Complete the implicit difference scheme \eqref{Maxwell4} with the
discrete analogues of the equations of motion of macroparticles:
\begin{equation}
\label{EquationOfMotion}
\begin{split}
&\frac{p_{zj}^{s+1}-p_{zj}^{s}}{\Delta t}=2Q_j\text{Re}\big(E_{0j}^{s+1/2}e^{ik_zz_j^{s+1/2}-i\omega(t+dt/2)}\big),p_j^{s+1/2}=\frac{p_j^{s+1}+p_j^{s}}{2},\\
&\frac{z_{j}^{s+1}-z_{j}^{s}}{\Delta t}=v_{zj}^{s+1/2},v_{zj}^{s+1/2}=\frac{p_{zj}^{s+1/2}/M_j}{\sqrt{1+(p_{zj}^{s+1/2}/M_jc)^2}}.\\
\end{split}
\end{equation}
The field $E_{j}^{s+1/2}$ at  particles' locations can be
found by means of linear interpolation from surrounding nodes:
\begin{equation}
 \label{Field}
\begin{split}
&E_{0j}^{s+1/2}=\frac{z_{n+1}-z_j^{s+1/2}}{\Delta z}E_{0n}^{s+1/2}+\frac{z_j^{s+1/2}-z_n}{\Delta z}E_{0n+1}^{s+1/2},\\
&z_n\le z_j^{s+1/2}<z_{n+1}.\\
\end{split}
\end{equation}

Injection and extraction of particles are performed as follows:
during every time step, we inject  $\Delta N$ number of particles,
whose coordinates are
\begin{equation*}
 z=-\Delta z\Big(1+\frac{i}{\Delta N}\Big),i=0,1,...,\Delta N-1.
\end{equation*}

Assume that each particle has a momentum $p_{z0}$ and a charge
$Q=j_{av}\Delta t+j_{rand}\Delta t$. Here $j_{av}$ is the average
current density  and $j_{rand}$ is a random quantity. This
quantity appears due to  shot noise of charged particles, which in
high power devices can achieve quite large values, because the
electron flow consists of ectons \cite{Abubakirov2009}. Since the
parametric radiation is narrow-band by nature, and the current
density spectrum $j_{rand}$ is a slowly varying function of
frequency $\omega$, one can replace the real current density
$j_{rand}$ by a random quantity uniformly distributed in the
domain $(-j_{rand0},j_{rand0})$. The level of noise is determined
by the fluctuation current density $j_{rand0}(\omega)$. By analogy
with the case of fluctuations of current, particle distribution
over momenta can be introduced into the system. Particles are
extracted from the system after they reach the coordinate $z_j\ge
L+\Delta z$.

\end{document}